\documentclass[11pt,a4paper]{article}
\usepackage[utf8]{inputenc}
\usepackage[T1]{fontenc}
\usepackage{amsmath}
\usepackage{amsfonts}
\usepackage{amssymb}
\usepackage{graphicx, url}
\usepackage[left=1.50cm, right=1.50cm, top=2.00cm, bottom=2.00cm]{geometry}

\usepackage{txfonts,comment}
\usepackage{comment,adjustbox}
\usepackage{tikz, circuitikz}
\usepackage{longtable, authblk}

\newcommand{\eh}{\textsc{Eye2Heart}~}
\numberwithin{equation}{section}

\title{\eh: a validated lumped-parameter model bridging cardiovascular and ocular dynamics}

\author[1,*]{Lorenzo Sala}
\author[2]{Mohamed Zaid}
\author[3]{Faith Hughes}
\author[4]{Marcela Szopos}
\author[5]{Virginia H. Huxley}
\author[6]{Alon Harris}
\author[3]{Giovanna Guidoboni}
\author[7]{Sergey Lapin}

\footnotesize
\affil[1]{Université Paris-Saclay, INRAE, MaIAGE, 78350, Jouy-en-Josas, France}
\affil[2]{Foresite Healthcare LLC, Maryland Heights, Missouri 63043, USA}
\affil[3]{Electrical and Computer Engineering, University of Maine, Maine, USA}
\affil[4]{Université Paris Cité, CNRS, MAP5, F-75006 Paris, France}
\affil[5]{Department of Medical Pharmacology and Physiology, School of Medicine, University of Missouri, Columbia, MO, United States}
\affil[6]{Department of Ophthalmology, Icahn School of Medicine at Mount Sinai, New York City, NY, 10029, USA}
\affil[7]{Department of Mathematics and Statistics, Program in Data Analytics, Washington State University Everett, Everett (WA), USA \vspace{2mm}}
\affil[*]{Corresponding author: \texttt{lorenzo.sala@inrae.fr}}

\date{}

\begin{document}

		\maketitle
		
		\vspace{-15mm}
		
		\begin{abstract}
			The cardiovascular and ocular systems are intricately connected, with hemodynamic interactions playing a crucial role in both physiological regulation and pathological conditions. However, existing models often treat these systems separately, limiting the understanding of their interdependence. 
			In this study, we present the \eh model, a novel closed-loop mathematical framework that integrates cardiovascular and ocular dynamics. 
			Using an electrical-hydraulic analogy, the model describes the interactions between the heart and retinal circulation through a system of ordinary differential equations. 
			The model is validated against clinical and experimental data, demonstrating its ability to reproduce key cardiovascular parameters (\textit{e.g.}, stroke volume, cardiac output) and ocular hemodynamics (\textit{e.g.}, retinal blood flow). 
			Additionally, we explore \textit{in silico} the effects of intraocular pressure (IOP) and left ventricular compliance on both local ocular and global systemic circulation, revealing critical dependencies between cardiovascular and ocular health. The results highlight the model’s potential for studying cardiovascular diseases with ocular manifestations, paving the way for patient-specific data integration and expanding applications in personalized medicine.		
		\end{abstract}
		
		\paragraph{\textbf{Keywords.}}
		Cardiovascular system, ocular hemodynamics, cardiovascular-ocular coupling, lumped-parameter model, ordinary differential equations, in silico experimentation, left ventricular compliance, intraocular pressure 
		
		\normalsize

		\section{Introduction}
		\label{sec:intro}
		
		The intricate relationship between the cardiovascular and ocular systems is essential for understanding a range of physiological and pathological processes \cite{flammer2013eye,HANSSEN2022101095}. Although both systems are crucial for maintaining systemic balance, their interdependence has not been thoroughly explored in current research. The cardiovascular system regulates systemic blood flow, which includes ocular perfusion, while intraocular pressure (IOP) and retinal blood flow are critical for maintaining ocular health.
		Studies have examined retinal vascular characteristics as indicators of cardiovascular health \cite{michelson1979retinal,tedeschietal2005}, and changes in cardiovascular status are known to affect retinal microvasculature \cite{abdin2024cardio}.
		Recent advancements, such as automated retinal photography and AI-based cardiovascular disease (CVD) risk assessment systems, have shown promise for non-invasive CVD risk evaluations, presenting a more streamlined alternative to traditional clinical methods \cite{hu2025real}. Additionally, the standardization and clinical use of retinal imaging biomarkers for CVD are becoming increasingly important in the effort to incorporate ocular data into cardiovascular risk assessments \cite{chew2025standardization}.
		
		Advances in AI-driven ocular image analysis have significantly contributed to CVD prediction, including the identification of risk factors and the potential replacement of conventional biomarkers \cite{huang2024ai}. 
		However, existing models often treat the cardiovascular and ocular systems as separate entities, concentrating on either cardiac and vascular dynamics or isolated ocular physiology.
		This compartmentalized approach has led to a limited understanding of how the two systems interact. A more integrated view is necessary to capture their physiological interdependence.
		Current cardiovascular models typically do not account for the unique hemodynamic requirements of ocular perfusion, while ocular models frequently overlook the dynamic effects of cardiac function. The lack of a unified modeling framework has constrained the potential for accurate simulations of cardiovascular-ocular interactions across various physiological states.
		A recent study by Caddy and co-authors \cite{caddy2024modelling} attempted to model large-scale arterial hemodynamics from the heart to the eye under simulated microgravity conditions. Although this research represents a significant step in connecting cardiovascular and ocular dynamics, it primarily focuses on arterial circulation without incorporating closed-loop feedback mechanisms, which are vital for fully understanding the complex interactions between these systems.
		
		Previous work from our research group has developed theoretical models of both cardiovascular \cite{guidoboni2019cardiovascular} and retinal networks \cite{guidoboni2014intraocular}, validated against clinical and experimental data.
		However, these models have remained isolated from each other. To address this limitation, we propose the development of a novel coupled \eh model. This model will serve as a virtual laboratory to investigate the integrated dynamics of cardiovascular and retinal blood circulation.
		The specific objectives of this research are twofold: (i) to develop and validate a closed-loop heart-eye model that integrates both cardiovascular and ocular dynamics, and (ii) to simulate clinically relevant scenarios, such as variations in IOP or changes in cardiac elastance, to explore system behavior under diverse conditions. Through these efforts, we aim to provide a more comprehensive understanding of cardiovascular and ocular health, leading to new insights for clinical applications and therapeutic interventions.
		
		The remainder of the paper is as follows: Section 2 presents the mathematical framework, the assumptions underlying the \eh model, and the value of the parameter employed in the model for baseline simulations. 
		Section 3 first details the validation process, comparing our baseline results with experimental and clinical data, and then explores key simulation scenarios, including variations in IOP and heart elastance. Finally, Section 4 discusses the implications of our findings, the limitations of the study, and directions for future research.
		
		\section{Model and methods}
		\label{sec:model}
		This work presents a novel closed-loop mathematical model designed to capture the intricate interplay between the ocular and cardiovascular circulation systems. A closed-loop model is a mathematical framework that encapsulates the interdependent feedback mechanisms between interconnected systems, ensuring that the output from one system serves as an input to the other, thus forming a continuous cycle. This approach aims to mimic the physiological interactions between the eye and the heart, where variations in one system can directly influence the other.
		
		The proposed coupling is based on mathematical equations derived from fundamental physiological principles, enabling an accurate simulation of this bidirectional relationship. In this section, we describe the development, implementation, and parameterization of the original closed-loop eye-heart model we designed, emphasizing its potential to provide novel insights into the integrated function of these critical systems. 
		
		\subsection{Closed-loop \eh model} \label{sec:closed_eyeheartmodel}
		
\begin{figure}[h!]
			\begin{center}
\begin{adjustbox}{scale=0.3}
\begin{circuitikz}[scale=0.95]
\large

\draw (2,-14) to [resistor, l^=$R_{aorta2eye,1}$] (2,-18);
\draw (4,-12) to [cute inductor, l=$L_{aorta2eye}$, i_<=\textcolor{red}{$Q_{aorta2eye}$}] (2,-14);
\draw (4,-12) to [short,-*, red, l=$P_{aorta2eye,1}$] (4,-12);
\draw (4,-12) to [capacitor, l=$C_{aorta2eye}$] (6,-14);
\draw (4,-7) to [resistor, l^=$R_{aorta2eye,2}$] (4,-12);
\draw (4,-7) to [short,-*, l_=$P_{aorta2eye,2}$] (4,-7);
\draw (6,-14) to (6,-14) node[ground]{};
\draw (4,6) to [short] (4,-7);
\draw (4,6) to [short] (-14,24);
\draw (-14,24) to [short] (-15,23);
\draw (-15,23) to [resistor, l^=$R_{eye,1}$] (-18.5,19.5);
\draw (-18.5,19.5) to [short,-*, red, l=$P_{eye}$] (-18.5,19.5);
\draw (-18.5,19.5) to [capacitor, l_={$C_{eye}$}] (-17.5,18.5);
\draw (-17.5,18.5) to (-17.5,18.5) node[ground, rotate=45]{};
\draw (-18.5,19.5) to [resistor, l^=$R_{eye,2}$] (-22,16);
\draw (-22,16) to [short] (-23,15);
\draw (-23,15) to [short] (-4,-4);
\draw (-4,-4) to [short] (-4,-7);
\draw (2,-5) to [resistor, l^=$R_{cra,in,1}$] (4,-7);
\draw (2,-5) to [capacitor, l_=$C_{cra,in}$] (1,-7);
\draw (2,-5) to [short,-*, red, l=$P_{cra,in}$] (2,-5);
\draw (2,-2) to [resistor, l^=$R_{cra,in,2}$] (2,-5);
\draw (1,-7) to (1,-7) node[ground]{};
\draw (-1,3) to [short,*-*] (2,-2);
\draw (-1,3) to [R, l_=$R_{cra,1a}$] (-3,5);
\draw (-3,5) to [short,-*, red, l^=$P_{cra,1}$] (-3,5);
\draw (-3,5) to [capacitor, l=$C_{cra,1}$] (-1,7);
\draw (-1,7) to (-1,7) node[rotate=135, ground]{};
\draw (-3,5) to [R, l_=$R_{cra,1b}$] (-5,7);
\draw (-5,7) to [R, l_=$R_{cra,2a}$] (-7,9);
\draw (-7,9) to [short,-*, l^=$P_{cra,2}$] (-7,9);
\draw (-7,9) to [R, l_=$R_{cra,2b}$] (-9,11);
\draw (-9,11) to [short] (-9,17);
\draw (-9,17) to [vR, l_=$R_{art,1}$] (-11,19);
\draw (-11,19) to [short,-*, red, l_=$P_{art}$] (-11,19);
\draw (-11,19) to [capacitor, l=$C_{art}$] (-13,17);
\draw (-13,17) to (-13,17) node[rotate=-45,ground]{};
\draw (-11,19) to [vR, l_=$R_{art,2}$] (-13,21);
\draw (-13,21) to [short] (-14,20);
\draw (-14,20) to [R, l_=$R_{cap,1}$] (-16,18);
\draw (-19,15) to [R, l^=$R_{cap,2}$] (-17,17);
\draw (-16.5,17.5) to [short,-*, l_=$P_{cap}$] (-16.5,17.5);
\draw (-17,17) to [short] (-16,18);
\draw (-20,14) to [short] (-19,15);
\draw (-18,12) to [vR, l^=$R_{ven,1}$] (-20,14);
\draw (-16,14) to (-16,14) node[ground, rotate=135]{};
\draw (-18,12) to [capacitor, l_=$C_{ven}$] (-16,14);
\draw (-18,12) to [short,-*, l^=$X_{15}$] (-18,12);
\draw (-16,10) to [vR, l^=$R_{ven,2}$] (-18,12);
\draw (-10,10) to [short] (-16,10);
\draw (-8,8) to [vR, l^=$R_{crv,1a}$] (-10,10);
\draw (-6,6) to [vR, l^=$R_{crv,1b}$] (-8,8);
\draw (-8,8) to [short,-*, l=$P_{crv,1}$] (-8,8);
\draw (-4,4) to [R, l^=$R_{crv,2a}$] (-6,6);
\draw (-6,2) to (-6,2) node[ground, rotate=-45]{};
\draw (-4,4) to [capacitor, l_=$C_{crv,2}$] (-6,2);
\draw (-4,4) to [short,-*, red, l_=$P_{crv,2}$] (-4,4);
\draw (-2,2) to [R, l^=$R_{crv,2b}$] (-4,4);

\draw (-2,2) to [short,*-] (-2,-2);
\draw (-2,-2) to [resistor, l^=$R_{crv,out,1}$] (-2,-5);
\draw (-2,-5) to [short,-*, red, l^=$P_{crv,out}$] (-2,-5);
\draw (-2,-5) to [capacitor, l^=$C_{crv,out}$] (-1,-7);
\draw (-1,-7) to (-1,-7) node[ground]{};
\draw (-2,-5) to [resistor, l^=$R_{crv,out,2}$] (-4,-7);
\draw (-4,-7) to [resistor, l^=$R_{eye2vc,1}$] (-4,-12);
\draw (-4,-7) to [short,-*, red, l^=$P_{eye2vc,1}$] (-4,-7);
\draw (-4,-12) to [capacitor, l_=$C_{eye2vc}$] (-6,-14);
\draw (-6,-14) to (-6,-14) node[ground]{};
\draw (-4,-12) to [short,-*, red, l_=$P_{eye2vc,2}$] (-4,-12);
\draw (-4,-12) to [cute inductor, l_=$L_{eye2vc}$, i=\textcolor{red}{$Q_{eye2vc}$}] (-2,-14);
\draw (-2,-14) to [resistor, l^=$R_{eye2vc,2}$] (-2,-18);

\draw (6,-22) to [short,-*, red, l^=$P_{aorta}$] (6,-22);
\draw (2,-18) to [short] (6,-22);
\draw (6,-22) to [resistor, l^=$R_{aorta,1}$] (5,-24);
\draw (5,-24) to [normal open switch, l=$S_{aortic}$] (4,-26);
\draw (4,-26) to [resistor, l^=$R_{LV}$] (2,-26);
\draw (2,-26) to [normal open switch,l =$S_{pv}$] (2,-24);
\draw (2,-24) to [resistor, l^=$R_{pv}$] (2,-22);
\draw (2,-26) to [variable capacitor, l=$E_{LV}$] (1,-28);
\draw (1,-27) node[left] {\textcolor{red}{$V_{LV}$}};
\draw (1,-28) to [american voltage source, l={$U_{LV}$}] (1,-29);
\draw (1,-29) to (1,-29) node[ground]{};
\draw (-6,-22) to [short,-*, red, l_=$P_{vc}$] (-6,-22);
\draw (-2,-18) to [short] (-6,-22);
\draw (-6,-22) to [resistor, l^=$R_{vc}$] (-5,-24);
\draw (-5,-24) to [normal open switch, l=$S_{vc}$] (-4,-26);
\draw (-4,-26) to [resistor, l^=$R_{RV}$] (-2,-26);
\draw (-2,-26) to [normal open switch, l_=$S_{pa}$] (-2,-24);
\draw (-2,-24) to [resistor, l_=$R_{pa}$] (-2,-22);
\draw (-2,-26) to [variable capacitor, l_=$E_{RV}$] (-1,-28);
\draw (-1,-28) to [american voltage source, l_=$U_{RV}$] (-1,-29);
\draw (-1,-29) to (-1,-29) node[ground]{};
\draw (-1,-27) node[right] {\textcolor{red}{$V_{RV}$}};

\draw (6,-22) to [short] (8,-22);
\draw (8,-22) to [short] (8,-40);
\draw (8,-40) to [short] (6,-40);
\draw (7,-40) to [resistor, l_=$R_{aorta,2}$] (4,-40);
\draw (7,-40) to [capacitor, l=$C_{aorta}$] (7,-42);
\draw (6,-42) to (6,-42) node[ground]{};
\draw (4,-40) to [cute inductor, l^=$L_{aorta}$, i_=\textcolor{red}{$Q_{aorta}$}] (1,-40);
\draw (1,-40) to [short] (-1,-40);
\draw (0.5,-40) to [short,-*, red, l_=$P_{body}$] (0,-40);
\draw (0,-40) to [capacitor, l_=$C_{body}$] (0,-42);
\draw (0,-42) to (0,-42) node[ground]{};
\draw (-7,-40) to [cute inductor, l_=$L_{body}$, i<=\textcolor{red}{$Q_{body}$}] (-4,-40);
\draw (-7,-40) to [capacitor, l_=$C_{vc}$] (-7,-42);
\draw (-6,-42) to (-6,-42) node[ground]{};
\draw (-4,-40) to [resistor, l^=$R_{body}$] (-1,-40);
\draw (-6,-22) to [short] (-8,-22);
\draw (-8,-22) to [short] (-8,-40);
\draw (-8,-40) to [short] (-6,-40);
\draw (6,-32) to [cute inductor, l^=$L_{pv}$, i<_=\textcolor{red}{$Q_{pv}$}] (4,-32);
\draw (6,-32) to [short,-*, red, l_=$P_{pv}$] (6,-32);
\draw (6,-32) to [capacitor, l_=$C_{pv}$] (6,-34);
\draw (6,-34) to (6,-34) node[ground]{};
\draw (4,-32) to [resistor, l^=$R_{lungs,2}$] (1,-32);
\draw (1,-32) to [short] (-1,-32);
\draw (1,-32) to [short,-*, red, l_=$P_{lungs}$] (-0,-32);
\draw (0,-32) to [capacitor, l_=$C_{lungs}$] (0,-34);
\draw (0,-34) to (0,-34) node[ground]{};
\draw (-6,-32) to [resistor, l^=$R_{lungs,1}$] (-4,-32);
\draw (-6,-31) to [short,-*, red, l_=$P_{pa}$] (-6,-32);
\draw (-6,-32) to [capacitor, l_=$C_{pa}$] (-6,-34);
\draw (-6,-34) to (-6,-34) node[ground]{};
\draw (-4,-32) to [cute inductor, l_=$L_{lungs}$, i=\textcolor{red}{$Q_{lungs}$}] (-1,-32);
\draw (6,-32) to [short] (6,-25);
\draw (6,-25) to [short] (5.5,-24.5);
\draw (5.5,-24.5) to[crossing] (4.5,-23.5);
\draw (4.5,-23.5) to [short] (2,-22);
\draw (-6,-32) to [short] (-6,-25);
\draw (-6,-25) to [short] (-5.5,-24.5);
\draw (-5.5,-24.5) to[crossing] (-4.5,-23.5);
\draw (-4.5,-23.5) to [short] (-2,-22);

\end{circuitikz}
\end{adjustbox}
\end{center}
			\caption{Cardiovascular model schematic of the \eh closed loop circuit.} 
		\label{fig:eye2heart_scheme}
	\end{figure}
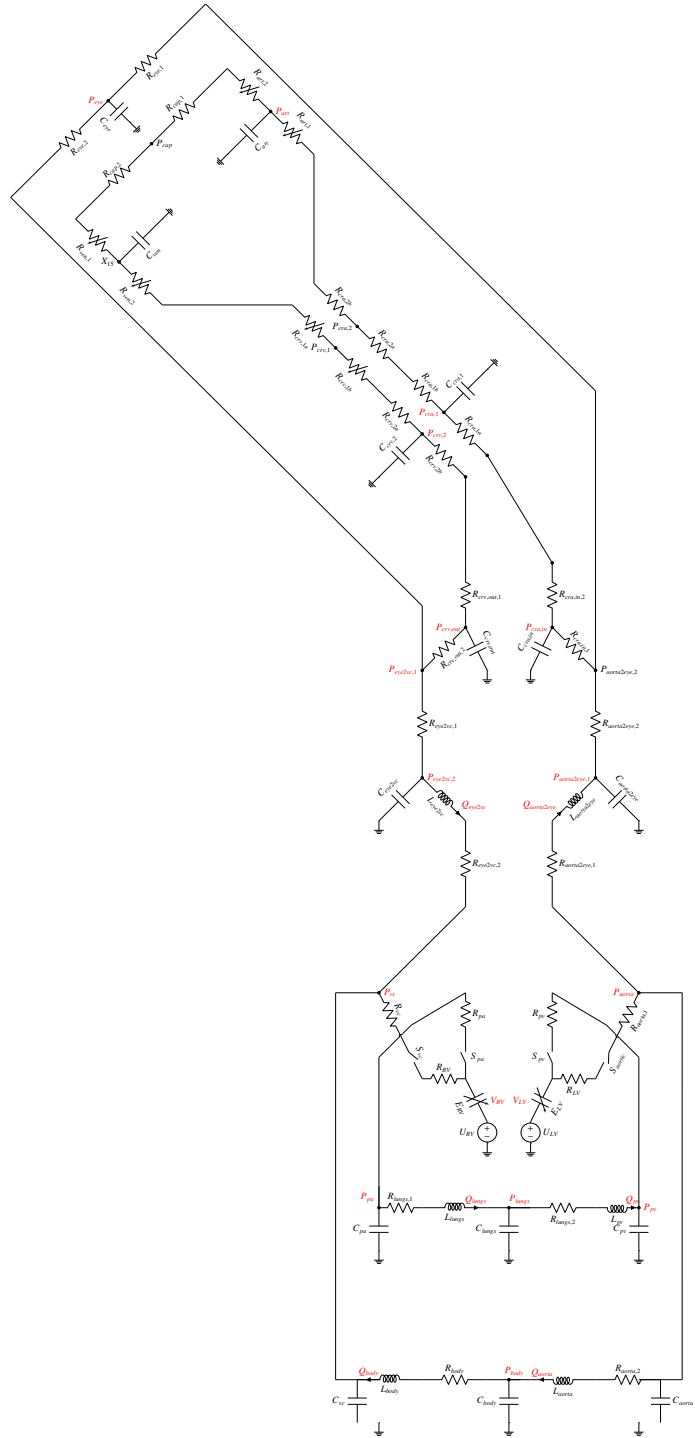

		Figure ~\ref{fig:eye2heart_scheme} illustrates the novel closed-loop Eye-Heart model, referred to as \eh hereafter.
		The system is built using an analogy between fluid flow in a hydraulic network and electrical current in a circuit. 
		In this framework, blood pressure, blood volume, and blood flow are represented by electric potential, electric charge, and electric current, respectively. Cardiovascular compartments, including blood vessels and heart ventricles, are modeled as combinations of resistances (R), representing the opposition to flow that blood is experiencing when flowing through a compartment, and compliances (C), representing the ability of that compartment to deform and store volume. 
		By writing Kirchhoﬀ laws for the nodes (conservation of current/ﬂow rate) and for closed circuits (conservation of the voltage/pressure difference), the resulting mathematical model is a system of 23 ordinary differential equations (ODEs) that capture the dynamic interactions between the ocular and the cardiovascular circulatory systems. This structure provides a robust foundation for analyzing and modeling the interconnected dynamics of these compartments. 
		For clarity, the description is divided into three parts: the cardiovascular subsystem, the ocular subsystem, and the eye-heart coupling dynamics. 
		To enhance the physiological interpretability of the model, the state variables and parameters have been renamed from their original terminology, as used in previous works by Avanzolini et al. \cite{avanzolini1988cadcs} and Guidoboni et al. \cite{guidoboni2014intraocular}. These updates better reflect the physiological roles of the variables and parameters within the context of the integrated model.
		
\vspace{-2mm}
\subsubsection*{Cardiovascular System}
The cardiovascular system in this model builds upon the mathematical framework proposed by Avanzolini et al~\cite{avanzolini1988cadcs}, modified to account for its coupling with the upper circulation that supplies blood specifically to one eye. The aortic pressure ($P_{\mathrm{aorta}}$) and the right venous-atrial pressure ($P_{\mathrm{vc}}$) are now interfaced with an \textit{ad hoc} upper circulation module, designed to model blood flow to the ocular compartment and specifically to the retinal circulation.

To incorporate these dynamics, the original equations have been reformulated, ensuring seamless integration of the new blood circulation pathway into the system. A key parameter, $R_{\mathrm{body}}$, which represents the equivalent peripheral resistance of the body, has been updated to reflect the redistribution of blood volume toward the eye. In the original model, $R_{\mathrm{body}}$ was assigned a value of \( 6.75 \cdot 10^{-2} \; [\mathrm{mmHg} \cdot \mathrm{s} \cdot \mathrm{ml}^{-1}] \). In the revised eye-heart model, this parameter has been adjusted to \( 6.93 \cdot 10^{-2} \; [\mathrm{mmHg} \cdot \mathrm{s} \cdot \mathrm{ml}^{-1}] \), capturing the additional resistance introduced by the retinal blood flow dynamics. 

Importantly, all other aspects of the cardiovascular system remain unchanged from the original model proposed by Avanzolini et al. \cite{avanzolini1988cadcs}. 
These modifications focus solely on integrating the eye-specific circulation while preserving the original framework for the rest of the cardiovascular system. As such, the adopted strategy ensures consistency with the validated physiological dynamics described in the original model.

\vspace{-2mm}
\subsubsection*{Ocular System}
A detailed description of the retinal circulation model and mathematical equations can be found in Guidoboni et al. \cite{guidoboni2014intraocular, guidoboni2019cardiovascular}.  
This section provides a brief overview of the main features of the model.
In this framework, the vasculature is divided into five main compartments labeled accordingly: the central retinal artery (CRA), arterioles (art), capillaries (cap), venules (ven), and the central retinal vein (CRV).
Alphanumerical labels further distinguish between specific segments within each compartment. 
The blood flow within the retina is driven by a pressure difference between the inlet, $P_{cra,in}$, which is the blood pressure upstream of the CRA, and the outlet, $P_{crv,out}$, the pressure downstream of the CRV.
External pressures affect different parts of the retinal network: intraocular segments are exposed to intraocular pressure (IOP), while retrobulbar segments behind the eye experience retrolaminar tissue pressure (RLTp). This combination of resistances and pressures provides a comprehensive blood flow model through the retinal vasculature.

The eye-heart model retains the retinal circulation framework established in the foundational work by Guidoboni et al.~\cite{guidoboni2014intraocular}, ensuring consistency with physiological values derived from experimental data. The original description of retinal circulation and its parameters, including vascular resistance, compliance, and blood flow dynamics, remain unchanged. In this manner, the model predictions remain consistent with validated physiological observations of retinal hemodynamics.

Moreover, building upon the base model of the retinal circulation, we further introduce an additional parallel circuit, referred to as the eye branch, which represents the blood flow directed toward non-retinal structures.
Although the term "eye branch" can be seen a simplification, it is intended to encompass all ocular vascular beds outside the retina, such as the choroid and ciliary body.  This extension enables the model to capture the broader ocular circulation, accounting for the distinct hemodynamic properties and functional roles of these non-retinal structures. 
By incorporating the eye branch, the model provides a more comprehensive representation of blood flow within the eye and accounts for flow redistribution mechanisms under conditions of elevated external pressure.

\vspace{-2mm}
\subsubsection*{Eye-Heart Coupling}
The original eye-heart coupling component of the physiological model is designed to capture the dynamics of blood circulation between the heart and a single eye. This focused approach simplifies the systemic circulatory model by isolating the blood flow directed to the eye while incorporating the rest of the blood flow, including that towards the brain and other systemic tissues, into a separate branch of the body circulation. This separation enables a targeted analysis of ocular hemodynamics without compromising the systemic integrity of the model.

The direct connection between the heart and the eye is described using an aorta-to-eye and eye-to-vena cava equivalent circuit, represented by resistive, capacitive, and inductive elements. These elements collectively simulate the vascular resistance, compliance, and volumetric blood flow within the pathway from the aorta to the ocular circulation and back to the venous system. By calibrating the R, C, and L parameters, the model captures the distinctive hemodynamic properties of the eye, including its dependence on systemic blood pressure and flow rates originating from the heart.

This simplification enhances computational efficiency and allows detailed exploration of the interplay of cardiovascular and ocular systems. It provides a robust framework for investigating eye-specific circulatory phenomena, such as retinal blood flow regulation and pressure-induced vascular changes.

\vspace{-2mm}
\subsubsection*{Model equations}
\label{subsec:eqs}
The overall description results in the following set of ODEs, which models the coupling between the eye and heart dynamics. 
These equations describe the hemodynamics between various compartments, including the aorta, body, lungs, and the eye. 
The system of equations accounts for pressure and flow rates in the cardiovascular and ocular systems, considering factors such as resistance, compliance, and inductance in each compartment. 
We gather below the equations governing the overall dynamics of the coupled system:

\begin{align}  
\dfrac{dP_{aorta}}{dt} &= \dfrac{1}{C_{aorta}} \left(\dfrac{(P_{LV}-P_{aorta})S_{aorta}}{R_{LV}-R_{aorta,1}} -Q_{aorta2eye}-Q_{aorta} \right) \\
\dfrac{dQ_{aorta}}{dt} &= \dfrac{1}{L_{aorta}} (P_{aorta}-P_{body}-R_{aorta,2}Q_{aorta}) \\
\dfrac{dP_{body}}{dt} &= \dfrac{1}{C_{body}} (Q_{aorta}-Q_{body}) \\
\dfrac{dQ_{body}}{dt} &= \dfrac{1}{L_{body}} (P_{body}-P_{VC}-R_{body}Q_{body})\\
\dfrac{dP_{VC}}{dt} &= \dfrac{1}{C_{VC}} \left((Q_{body}+Q_{eye2VC})- \dfrac{(P_{VC}-P_{RV})S_{VC}}{R_{VC}} \right) \\
\dfrac{dV_{RV}}{dt} &= \dfrac{(P_{VC}-P_{PA})S_{VC}}{R_{VC}} - \dfrac{(P_{RV}-P_{PA})S_{PA}}{R_{RV}+R_{PA}} \\
\dfrac{dP_{PA}}{dt} &= \dfrac{1}{C_{PA}} \left(\dfrac{(P_{RV}-P_{PA})S_{PA}}{R_{RV}+R_{PA}}-Q_{lungs} \right) \\
\dfrac{dQ_{lungs}}{dt} &= \dfrac{1}{L_{lungs}} (P_{pa}-P_{lungs}-R_{lungs,1}Q_{lungs}) \\
\dfrac{dP_{lungs}}{dt} &= \dfrac{1}{C_{PA}} (Q_{lungs}-Q_{PV}) \\
\dfrac{dQ_{PV}}{dt} &= \dfrac{1}{L_{PV}} (P_{lungs}-P_{PV}-R_{lungs}Q_{PV}) \\ 
\dfrac{dP_{PV}}{dt} &= \dfrac{1}{C_{PV}} \left(\dfrac{(P_{PV}-P_{LV})S_{PV}}{R_{PV}}-Q_{PV}\right) \\ 
\dfrac{dV_{LV}}{dt} &= \dfrac{(P_{PV}-P_{LV})S_{PV}}{R_{PV}} - \dfrac{(P_{LV}-P_{aorta})S_{aorta}}{R_{LV}+R_{aorta,1}} \\
\dfrac{dQ_{aorta2eye}}{dt} &= \dfrac{1}{L_{aorta2eye}} (P_{aorta}-P_{aorta2eye}-R_{aorta2eye,1}Q_{aorta2eye}) \\
\dfrac{dP_{aorta2eye,1}}{dt} &= \dfrac{1}{C_{aorta2eye}} \left(Q_{aorta2eye} - \dfrac{P_{aorta2eye,1}-P_{aorta2eye,2}}{R_{aorta2eye,2}} \right) \\
\dfrac{dP_{eye}}{dt} &= \dfrac{1}{C_{eye}} \left(\dfrac{P_{aorta2eye,2}-P_{eye}}{R_{eye,1}}-\dfrac{P_{eye}-P_{eye2VC}}{R_{eye,2}} \right) \\
\dfrac{dP_{CRAin}}{dt} &= \dfrac{1}{C_{CRAin}} \left(\dfrac{P_{aorta2eye,2}-P_{CRAin}}{R_{CRAin,1}} - \dfrac{P_{CRAin}-P_{CRA,1}}{R_{CRAin,2}+R_{CRA,1a}} \right) \\
\dfrac{dP_{CRA,1}}{dt} &= \dfrac{1}{C_{CRA,1}}    \left(\dfrac{P_{CRAin}-P_{CRA,1}}       {R_{CRAin,2}+R_{CRA,1a}} - \dfrac{P_{CRA,1}-P_{art}}{R_{CRA,1b}+R_{CRA,2a}+R_{CRA,2b}+R_{art,1}} \right) \\ 
\dfrac{dP_{art}}{dt} &= \dfrac{1}{C_{art}} \left(\dfrac{P_{CRA,1}-P_{art}}{R_{CRA,1b}+R_{CRA,2a}+R_{CRA,2b}+R_{art,1}}-\dfrac{P_{art}-P_{ven}}{R_{art,2}+R_{cap,1}+R_{cap,2}+R_{ven,1}} \right) \\ 
\dfrac{dP_{ven}}{dt} &= \dfrac{1}{C_{ven}} \left(\dfrac{P_{art}-P_{ven}}{R_{art,1}+R_{cap,1}+R_{cap,2}+R_{ven,1}} - \dfrac{P_{ven}-P_{CRV,2}}{R_{ven,2}+R_{CRV,1a}+R_{CRV,1b}+R_{CRV,2a}} \right) \\ 
\dfrac{dP_{CRV,2}}{dt} &= \dfrac{1}{C_{CRV,1}} \left(\dfrac{P_{ven}-P_{CRV,2}}{R_{ven,2}+R_{CRV,1a}+R_{CRV,1b}+R_{CRV,2a}} -\dfrac{P_{CRV,2}-P_{CRVout}}{R_{CRV,2b}+R_{CRVout,1}} \right) \\
\dfrac{dP_{CRVout}}{dt} &= \dfrac{1}{C_{CRVout}} \left(\dfrac{P_{CRV,2}-P_{CRVout}}{R_{CRV,2b}+R_{CRVout,1}} - \dfrac{P_{CRVout}-P_{eye2VC,1}}{R_{CRVout,2}} \right) \\
\dfrac{dP_{eye2VC}}{dt} &= \dfrac{1}{C_{eye2VC}} \left(\dfrac{P_{eye2VC,1}-P_{eye2VC,2}}{R_{eye2VC,1}}-Q_{eye2VC} \right) \\
\dfrac{dQ_{eye2VC}}{dt} &= \dfrac{1}{L_{eye2VC}} (P_{eye2VC,2}-P_{vc}-R_{eye2VC,2}Q_{eye2VC})
\end{align}

\subsection{Model parameterization and numerical methods} \label{subsec:param}

In this section, we outline the key physiological parameters used in the model, including their baseline values for the numerical simulations. 
We start by providing in Table~\ref{tab:variables_summary} a list of the model state variables, together with the associated initial conditions employed in the simulations.

Next, parameters are categorized into cardiovascular system, heart system, ocular hemodynamics, and cardiovascular-ocular connections and listed in Table~\ref{tab:parameters_summary}. 
These parameters influence the dynamics of blood flow and pressure, as well as the resistance and compliance in different vascular compartments. 
The initial conditions and parameter values in this study have been calibrated to ensure baseline flow and pressure consistency in the ocular system. 
To establish reliable estimates for vascular resistances, preliminary computations were performed under steady-state conditions.

\footnotesize
\begin{longtable}{ccccc}
 \caption{Summary of model state variables. \\ CRA=Central Retinal Artery; CRV=Central Retinal Vein.}\label{tab:variables_summary} \\
 \hline \hline
    \textsc{Variable} & \textsc{Description} & \textsc{Initial condition} & Units & \textsc{Reference} \\
        \hline\hline
    Cardiovascular & & & & \\ 
    $P_{aorta}$ & aortic pressure & $90.1$ & mmHg & \cite{avanzolini1988cadcs} \\
    $P_{body}$ & body pressure & $70.5$ & mmHg & \cite{avanzolini1988cadcs} \\
    $P_{VC}$ & vena cava pressure & $3.32$ & mmHg & \cite{avanzolini1988cadcs} \\
    $P_{PA}$ & pulmonary artery pressure & $13.4$ & mmHg & \cite{avanzolini1988cadcs} \\
    $P_{lungs}$ & lungs pressure & $13.3$ & mmHg & \cite{avanzolini1988cadcs} \\
    $P_{PV}$ & pulmonary vein pressure & $11.2$ & mmHg & \cite{avanzolini1988cadcs} \\
    $Q_{aorta}$ & aortic flow rate & $8.89$ & ml/s & \cite{avanzolini1988cadcs} \\
    $Q_{body}$ & body flow rate & $67.3$ & ml/s & \cite{avanzolini1988cadcs} \\ 
    $Q_{lungs}$ & lungs flow rate & $0.78$ & ml/s & \cite{avanzolini1988cadcs} \\
    $Q_{PV}$ & pulmonary vein flow rate & $23.8$ & ml/s & \cite{avanzolini1988cadcs} \\ 
    $V_{RV}$ & vena cava volume & $105$ & ml & \cite{avanzolini1988cadcs} \\
    $V_{LV}$ & left ventricle volume & $112$ & ml & \cite{avanzolini1988cadcs} \\
    Eye-Heart Coupling & & & & \\ 
    $P_{aorta2eye,1}$ & aorta-to-eye pressure & $80.25$ & mmHg & This work \\
    $P_{CRAin}$ & CRA pressure & $70.2$ & mmHg & This work \\
    $P_{CRVout}$ & CRV pressure & $8.57$ & mmHg & This work \\
    $P_{eye}$ & eye pressure & $65.5$ & mmHg & This work \\
    $P_{eye2vc,2}$ & eye-to-vena cava pressure & $4.52$ & mmHg & This work \\
    $Q_{eye2vc}$ & eye-to-vena-cava flow rate & $0.15$ & ml/s & This work \\
    $Q_{aorta2eye}$ & aorta-to-eye flow rate & $0.15$ & ml/s & This work \\
   Ocular Circulation & & & & \\ 
   $P_{CRA,1}$ & CRA pressure & $43.5$ & mmHg & \cite{guidoboni2014intraocular} \\
   $P_{art}$ & arteriole pressure & $35.5$ & mmHg & \cite{guidoboni2014intraocular} \\
   $P_{ven}$ & venule pressure & $21.8$ & mmHg & \cite{guidoboni2014intraocular} \\
   $P_{CRV,2}$ & CRV pressure & $18.9$ & mmHg & \cite{guidoboni2014intraocular} \\
     \hline\hline
\end{longtable}
\normalsize

\footnotesize
\begin{longtable}{ccccc}
 \caption{Summary of model parameters. \\ CRA=Central Retinal Artery; CRV=Central Retinal Vein.} \label{tab:parameters_summary} \\
    \hline \hline
    \textsc{Symbol} & \textsc{Description} & \textsc{Value} & \textsc{Units} & \textsc{Reference} \\
        \hline\hline
    \multicolumn{4}{l}{\textit{Cardiovascular system}}\\
    $R_{aorta,1}$ & aortic resistance & $3.751 \cdot 10^{-3}$ & mmHg s / ml & \cite{avanzolini1988cadcs} \\
    $R_{aorta,2}$ & aortic resistance & $6.93\cdot 10^{-2}$ & mmHg s / ml & This work \\
    $R_{body}$ & body resistance & $1.0$ & mmHg s / ml & \cite{avanzolini1988cadcs} \\
    $R_{vc}$ & vena cava resistance & $3.751\cdot 10^{-3}$ & mmHg s / ml & \cite{avanzolini1988cadcs} \\
    $R_{pa}$ & pulmonary artery resistance & $3.751\cdot 10^{-3}$ & mmHg s / ml & \cite{avanzolini1988cadcs} \\
    $R_{lungs,1}$ & lungs resistance & $3.376\cdot 10^{-2}$ &  mmHg s / ml & \cite{avanzolini1988cadcs} \\
    $R_{lungs,2}$ & lungs resistance & 0.1013 &  mmHg s / ml & \cite{avanzolini1988cadcs} \\
    $R_{pv}$ & pulmonary vein resistance & $3.751\cdot 10^{-3}$ &  mmHg s / ml & \cite{avanzolini1988cadcs} \\
    $C_{aorta}$ & aortic compliance & $0.22$ & ml / mmHg & \cite{avanzolini1988cadcs} \\
    $C_{body}$ & body compliance & $1.46$ & ml / mmHg& \cite{avanzolini1988cadcs} \\
    $C_{vc}$ & vena cava compliance & $20.0$ & ml / mmHg & \cite{avanzolini1988cadcs} \\
    $C_{pa}$ & pulmonary artery compliance & $9.0\cdot 10^{-2}$ & ml / mmHg & \cite{avanzolini1988cadcs} \\
    $C_{lungs}$ & lungs capacitance & 2.67 & ml / mmHg & \cite{avanzolini1988cadcs} \\
    $C_{pv}$ & pulmonary vein capacitance & $46.7$ & ml / mmHg & \cite{avanzolini1988cadcs} \\
    $L_{aorta}$ & aortic fluid inertance & $8.25\cdot 10^{-4}$ & mmHg s$^2$ / ml & \cite{avanzolini1988cadcs} \\
    $L_{body}$ & body fluid inertance & $3.6\cdot 10^{-3}$ & mmHg s$^2$ / ml & \cite{avanzolini1988cadcs} \\
    $L_{lungs}$ & lungs fluid inertance & $7.5\cdot 10^{-4}$ & mmHg s$^2$ / ml & \cite{avanzolini1988cadcs} \\
    $L_{pv}$ & pulmonary vein fluid inertance & $3.08\cdot 10^{-3}$ & mmHg s$^2$ / ml & \cite{avanzolini1988cadcs} \\
    \noalign{\smallskip} \hline
    \multicolumn{4}{l}{\textit{Heart system}}\\
    $R_{LV}$ & left ventricle resistance & $8.0\cdot 10^{-3}$ &  mmHg s / ml & \cite{avanzolini1988cadcs} \\
    $R_{RV}$ & right ventricle resistance & $1.75\cdot 10^{-2}$ &  mmHg s / ml & \cite{avanzolini1988cadcs} \\
    $U_{L0}$ & left ventricle isovolumic pressure & 50.0 & mmHg & \cite{avanzolini1988cadcs} \\
    $E_{LD}$ & left ventricle diastolic elastance & 0.1 & mmHg / ml & \cite{avanzolini1988cadcs} \\
    $E_{LS}$ & left ventricle systolic elastance & 1.375 & mmHg / ml & \cite{avanzolini1988cadcs} \\
    $U_{R0}$ & right ventricle isovolumic pressure & 24.0 & mmHg & \cite{avanzolini1988cadcs} \\
    $E_{RD}$ & right ventricle diastolic elastance & $3.0\cdot 10^{-2}$ & mmHg / ml & \cite{avanzolini1988cadcs} \\
    $E_{RS}$ & right ventricle systolic elastance & 0.3288 & mmHg / ml & \cite{avanzolini1988cadcs} \\
    \noalign{\smallskip} \hline
    \multicolumn{4}{l}{\textit{Ocular hemodynamics}}\\
    IOP & intraocular pressure & 15 & mmHg \\
    $R_{cra,1a}$ & CRA resistance & $2.68\cdot 10^{4}$ &  mmHg s / ml & \cite{guidoboni2014intraocular} \\
    $R_{cra,1b}$  & CRA resistance & $4.3\cdot 10^{3}$ &  mmHg s / ml & \cite{guidoboni2014intraocular} \\
    $R_{art,1}$ & retinal arterioles resistance & $6.0\cdot 10^{3}$&  mmHg s / ml & \cite{guidoboni2014intraocular} \\
    $R_{art,2}$   & retinal arterioles resistance & $6.0\cdot 10^{3}$&  mmHg s / ml & \cite{guidoboni2014intraocular} \\
    $R_{cap,1}$  & retinal capillaries resistance & $5.68\cdot 10^{3}$&  mmHg s / ml & \cite{guidoboni2014intraocular} \\
    $R_{cap,2}$ & retinal capillaries resistance & $5.68\cdot 10^{3}$&  mmHg s / ml & \cite{guidoboni2014intraocular} \\
    $R_{crv,2a}$  & CRV resistance & $1.35\cdot 10^{3}$&  mmHg s / ml & \cite{guidoboni2014intraocular} \\
    $R_{crv,2b}$ & CRV resistance &   $22.09\cdot 10^{3}$&  mmHg s / ml & \cite{guidoboni2014intraocular} \\
    $K_{l_{cra,2a}}$ & CRA nonlinear resistance  parameter & 58.223 & [-] & \cite{guidoboni2014intraocular} \\
    $K_{l_{cra,2b}}$ & CRA nonlinear resistance  parameter & 58.223 & [-] & \cite{guidoboni2014intraocular} \\
    $K_{p_{cra,2a}}$  & CRA nonlinear resistance  parameter & 23.0894 & mmHg & \cite{guidoboni2014intraocular} \\
    $K_{p_{cra,2b}}$  & CRA nonlinear resistance  parameter & 23.0894 & mmHg & \cite{guidoboni2014intraocular} \\
    $k_{0_{cra,2a}}$  & CRA nonlinear resistance  parameter & 0.005115  & [-] & \cite{guidoboni2014intraocular} \\
    $k_{0_{cra,2b}}$  & CRA nonlinear resistance  parameter & 0.001023 & [-] & \cite{guidoboni2014intraocular} \\
    $K_{l_{crv,1a}}$ & CRV nonlinear resistance  parameter & $1.48425\cdot 10^{3}$ & [-] & \cite{guidoboni2014intraocular} \\
    $K_{l_{crv,1b}}$ & CRV nonlinear resistance  parameter & $1.48425\cdot 10^{3}$ & [-] & \cite{guidoboni2014intraocular} \\
    $K_{p_{crv,1a}}$ & CRV nonlinear resistance  parameter & 0.358774 & mmHg & \cite{guidoboni2014intraocular} \\
    $K_{p_{crv,1b}}$ & CRV nonlinear resistance  parameter & 0.358774 & mmHg & \cite{guidoboni2014intraocular} \\
    $k_{0_{crv,1a}}$ & CRV nonlinear resistance  parameter & 0.00324 & [-] & \cite{guidoboni2014intraocular} \\
    $k_{0_{crv,1b}}$ & CRV nonlinear resistance  parameter & 0.0162 & [-] & \cite{guidoboni2014intraocular} \\
    $K_{l_{ven,1}}$    &  retinal venules nonlinear resistance parameter & $1.2\cdot 10^{3}$ & [-] & \cite{guidoboni2014intraocular} \\
    $K_{l_{ven,2}}$    &  retinal venules nonlinear resistance parameter & $1.2\cdot 10^{3}$ & [-] & \cite{guidoboni2014intraocular} \\
    $K_{p_{ven,1}}$    &  retinal venules nonlinear resistance parameter & $0.0543$  & mmHg & \cite{guidoboni2014intraocular} \\
    $K_{p_{ven,2}}$    &  retinal venules nonlinear resistance parameter & $0.0543$  & mmHg & \cite{guidoboni2014intraocular} \\
    $k_{0_{ven,1}}$   &  retinal venules nonlinear resistance parameter & $2.8025\cdot 10^{-4}$ & [-] & \cite{guidoboni2014intraocular} \\
    $k_{0_{ven,2}}$   &  retinal venules nonlinear resistance parameter & $2.8025 \cdot 10^{-4}$ & [-] & \cite{guidoboni2014intraocular} \\
    $C_{cra,1}$ & CRA compliance & $7.22\cdot 10^{-7}$ & ml / mmHg & \cite{guidoboni2014intraocular} \\
    $C_{art}$ & retinal arterioles compliance & $7.53\cdot 10^{-7}$ & ml / mmHg & \cite{guidoboni2014intraocular} \\
    $C_{ven}$ & retinal venules compliance & $1.67\cdot 10^{-5}$ & ml / mmHg & \cite{guidoboni2014intraocular} \\
    $C_{crv,2}$ & CRV compliance & $1.07\cdot 10^{-5}$ & ml / mmHg & \cite{guidoboni2014intraocular} \\
    \noalign{\smallskip} \hline
    \multicolumn{4}{l}{\textit{Cardiovascular - ocular connection}}\\
    $R_{aorta2eye,1}$ & aorta-to-eye resistance & 55.062 &  mmHg s / ml & This work \\
    $R_{aorta2eye,2}$ & aorta-to-eye resistance & 55.062 &  mmHg s / ml & This work \\
    $R_{cra,in,1}$      &   pre-laminar CRA resistance & 5254.828 &  mmHg s / ml & This work \\
    $R_{cra,in,2}$      &   pre-laminar CRA resistance & 5254.828 &  mmHg s / ml & This work \\
    $R_{crv,out,1}$      &   pre-laminar CRV resistance & 16331.607 &  mmHg s / ml & This work \\
    $R_{crv,out,2}$      &   pre-laminar CRV resistance & 16331.607 &  mmHg s / ml & This work \\
    $R_{eye2vc,1}$ & aorta-to-vena cava resistance & 4.599&  mmHg s / ml & This work \\
    $R_{eye2vc,2}$ & aorta-to-vena cava resistance & 4.599&  mmHg s / ml & This work \\
    $R_{eye,1}$  & eye resistance & 268.777 &  mmHg s / ml & This work \\
    $R_{eye,2}$  & eye resistance & 268.777 &  mmHg s / ml & This work \\
    $C_{aorta2eye}$ & aorta-to-eye compliance & $1.66125\cdot 10^{-5}$ & ml / mmHg & This work \\
    $C_{{cra,in}}$ & pre-laminar CRA compliance & $1.72\cdot 10^{-5}$ & ml / mmHg & This work \\
    $C_{eye}$ & eye compliance & $3.6125\cdot 10^{-6}$ & ml / mmHg & This work \\
    $C_{crv,out}$  & pre-laminar CRV compliance & $1.6125\cdot 10^{-4}$ & ml / mmHg & This work \\
    $C_{eye2vc}$ & eye-to-vena cava compliance & $6.58\cdot 10^{-8}$ & ml / mmHg & This work \\
    $L_{aorta2eye}$ & aorta-to-eye fluid inertance & 0.0343 & mmHg s$^2$ / ml & This work \\
    $L_{eye2vc}$ & eye-to-vena cava fluid inertance & 0.0042 & mmHg s$^2$ / ml & This work \\
     \hline\hline
\end{longtable}
\normalsize

\noindent

\subsection{Numerical Solution of the \eh Closed-Loop Model}
The \eh mathematical model was implemented in MATLAB and solved using the stiff solver ODE15s \cite{shampine1999solving}, a variable-step, variable-order solver designed to efficiently solve stiff ordinary differential equations (ODEs). ODE15s was chosen to solve the system of ODEs due to its ability to efficiently handle stiffness, ensuring stable and accurate numerical solutions. The solver was configured with a relative tolerance of $10^{-13}$ and an absolute tolerance $10^{-5}$, ensuring high numerical precision. A fixed time step of 0.001 s was used throughout the simulations. To minimize numerical artifacts, transient effects at the beginning of the simulation were discarded.

		\section{Simulation results of the \eh model}
		\label{sec:results}

\subsection{Model validation}
\label{subsec:validation}

The model's cardiovascular component is validated through comparison with well-established physiological indicators. 
The following key parameters are computed and contrasted with clinical literature:
\begin{itemize}
	\item End-Diastolic Volume (EDV): Maximum ventricular volume during the cardiac cycle.
	\item End-Systolic Volume (ESV): Minimum ventricular volume during the cardiac cycle.
	\item Stroke Volume (SV): $\displaystyle SV = EDV - ESV$, representing the amount of blood ejected per beat.
	\item Cardiac Output (CO): $\displaystyle CO = HR \times \frac{SV}{1000}$, measuring total blood flow per minute, where $HR$ is heart rate.
	\item Ejection Fraction (EF): $\displaystyle EF = 100 \times \frac{SV}{EDV}$, quantifying ventricular efficiency as the percentage of blood ejected during each beat.
	\item End-Systolic Elastance (E$_{es}$): slope of the end-systolic pressure-volume relationship.
	\item Arterial Elastance (E$_a$): estimate of aortic input impedance.
	\item Central Systolic and Diastolic Pressures (SP/DP).
	\item Right Atrial Pressure (P$_{ra}$).
\end{itemize}
Table~\ref{tab:Heart_Values} presents a comparison between model predictions and clinical reference values, considering these quantitative indicators.

\begin{table}[h!]
	\centering
	\begin{tabular}{llcccc}
		\hline
		\textsc{Parameter} & \textsc{Unit} &  \multicolumn{2}{c}{\textsc{Clinical Ranges from literature}} & \multicolumn{2}{c}{\textsc{Present work} } \\[.03in]
		&& \emph{Left} & \emph{Right} & \emph{Left } & \emph{Right} \\
		&& \emph{Ventricle} & \emph{Ventricle} & \emph{Ventricle} & \emph{Ventricle} \\
		\hline\\
		End-Diastolic  & [ml] & 
		142 (102,183)~\cite{Maceira2006Left} &
		144 (98,190)~\cite{Maceira2006Right} &
		112.76 &
		115.25 \\
		Volume (EDV)& & 108 $\pm$27~\cite{sandstede2000} & 115 $\pm$ 31~\cite{sandstede2000} & & \\
		& & 109 $\pm$ 27~\cite{MAURER2007972} & 100 - 160~\cite{Chart} & & 
		\\[.08in]

		End-Systolic  & [ml] & 
		47 (27,68)~\cite{Maceira2006Left} &
		50 (22,78)~\cite{Maceira2006Right} &
		42.59 &
		43.81 \\
		Volume (ESV)& & 35 $\pm$13 ~\cite{sandstede2000} & 43 $\pm$ 19~\cite{sandstede2000} & & \\
		& & 30 $\pm$ 12~\cite{MAURER2007972} & 50-100~\cite{Chart} & & 
		\\[.08in]

		Stroke  & [ml/beat] & 
		95 (67, 123)~\cite{Maceira2006Left} &
		94 (64, 124)~\cite{Maceira2006Right} &
		70.18 &
		71.45 \\
		Volume (SV)& & 60 - 100~\cite{Chart}& 60-100~\cite{Chart} \\
		$ $ & $ $ & 81 $\pm$ 18~\cite{simone1997} \\
		& & 78 $\pm$ 20~\cite{MAURER2007972}
		\\[.08in]
		Cardiac & [l/min] & 
		4-8~\cite{Chart}  &
		4-8~\cite{Chart} &
		5.26 &
		5.36 \\
		Output (CO) & & 5.524 $\pm$ 1.488~\cite{simone1997} & \\
		& & 4.8 $\pm$ 1.3 ~\cite{MAURER2007972} 
		\\[.08in]
		Ejection & [\%] & 
		67 (58, 76)~\cite{Maceira2006Left}   &
		66 (54, 78)~\cite{Maceira2006Right}&
		62.32 &
		61.99 \\
		Fraction (EF) & & 72 $\pm$ 7~\cite{MAURER2007972} & 40 - 60~\cite{Chart} & & \\[.08in]
		End-Systolic & [mmHg/ml] & {1.74} \cite{redfield2005age} &$0.7 \pm 0.2$ \cite{singh2021sex} & $1.03$ & $0.32$ \\
		Elastance (E$_{es}$) & \\[.08in]
		Arterial & [mmHg/ml] & {1.2} \cite{redfield2005age} & $0.5 \pm 0.2$ \cite{singh2021sex} & $1.65$ & $0.52$ \\
		Elastance (E$_a$)  & \\[.08in]
		Central Systolic & [mmHg] & \multicolumn{2}{c}{$124.1 \pm 11.1$ \cite{sesso2000systolic}} & \multicolumn{2}{c}{$125.7$}\\
		Pressure (SP) & \\[.08in]
		Central Diastolic & [mmHg] & \multicolumn{2}{c}{$77.5 \pm 7.1$ \cite{sesso2000systolic}} & \multicolumn{2}{c}{$72.7$}\\
		Pressure (DP) & \\[.08in]
		Right Atrial & [mmHg] & \multicolumn{2}{c}{$3 \pm 2$ \cite{LOCK2006213}} & \multicolumn{2}{c}{$3.78$}\\
		Pressure (P$_{ra}$) & 
		\\[.05in]
		\hline \\[.08in]
	\end{tabular}
	\caption{Cardiovascular physiology indicators, left and right ventricles: comparison between clinical ranges from literature and simulation results from the present work.}
	\label{tab:Heart_Values}
\end{table}

\paragraph{The model predictions for key cardiovascular biomarkers,} including ventricular volumes, stroke volume, cardiac output, and ejection fraction, align well with reported clinical values. Notably, EDV and ESV for both left and right ventricles fall within the expected clinical ranges, with only slight underestimations compared to some references. 
This discrepancy may be attributed to model simplifications in ventricular compliance or assumptions on pressure-volume relationships.

SV is well captured, remaining within the accepted physiological range, reinforcing the model's ability to replicate fundamental cardiac dynamics. Similarly, the predicted CO falls within the expected $4-8$ L/min range, demonstrating the model's ability to effectively integrate heart rate and stroke volume.

The EF, a key indicator of ventricular function, is slightly lower than some clinical values but remains within a reasonable physiological range. This could suggest that the model, while accurately capturing systolic and diastolic phases, may benefit from refinements in contractility representation or vascular resistance calibration. Nonetheless, the overall agreement with literature values indicates a robust framework for modeling cardiac function.

In terms of E$_{es}$, the model predicts a value of of $1.03$ mmHg/ml for the left ventricle and $0.32$ mmHg/ml for the right ventricle.
These values are slightly higher than those reported in \cite{singh2021sex} and \cite{redfield2005age}, respectively. This discrepancy may stem from the model’s simplified assumptions regarding the pressure-volume relationship and elastance dynamics. Furthermore, the model does not account for individual variability, particularly for the left ventricle, which could contribute to the observed differences.

Similarly, for E$_a$, the model predicts slightly higher values for the left ventricle with respect to the clinical value, which, however, might be due to individual variabilities not reported in \cite{redfield2005age}.
The prediction for the right ventricle aligns well with the clinical value, suggesting that the model accurately captures arterial elastance for this chamber. 

Regarding SP and DP, the model predictions of $125.7$ mmHg and $72.7$ mmHg, respectively, agree with the clinical values.

Finally, for $P_{ra}$, the model prediction of $3.78$ mmHg is in good agreement with the clinical range of $3 \pm 2$ mmHg \cite{LOCK2006213}, indicating that the model accurately captures this biomarker of the venous pressure system.

\begin{figure}[ht]
	\centering
	\begin{minipage}{0.49\linewidth}
		\centering
		\includegraphics[width=\linewidth]{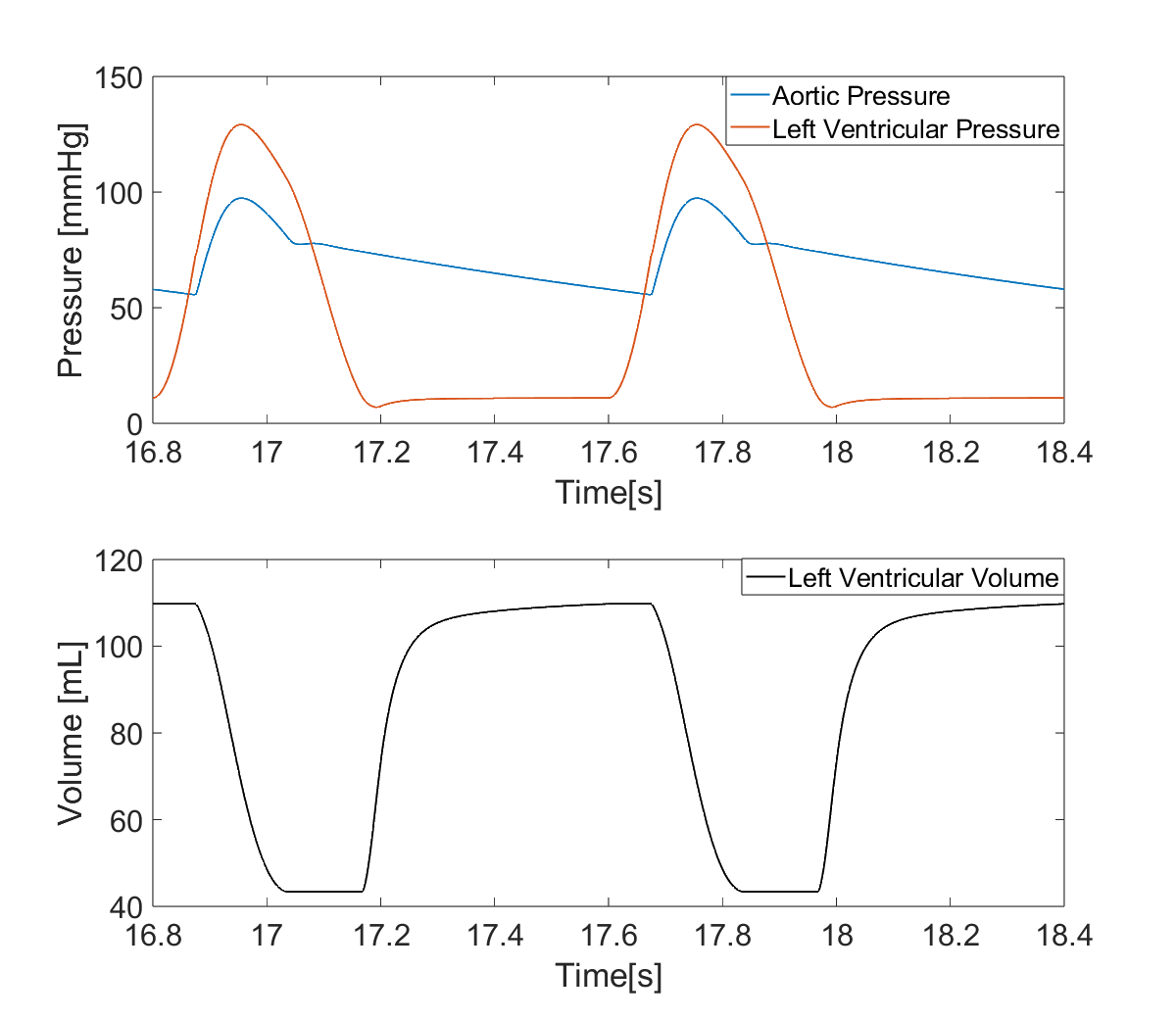}
		\caption{Wiggers diagram.}
		\label{fig:validation:wiggers}
	\end{minipage} \hfill
	\begin{minipage}{0.49\linewidth}
		\includegraphics[width=\linewidth]{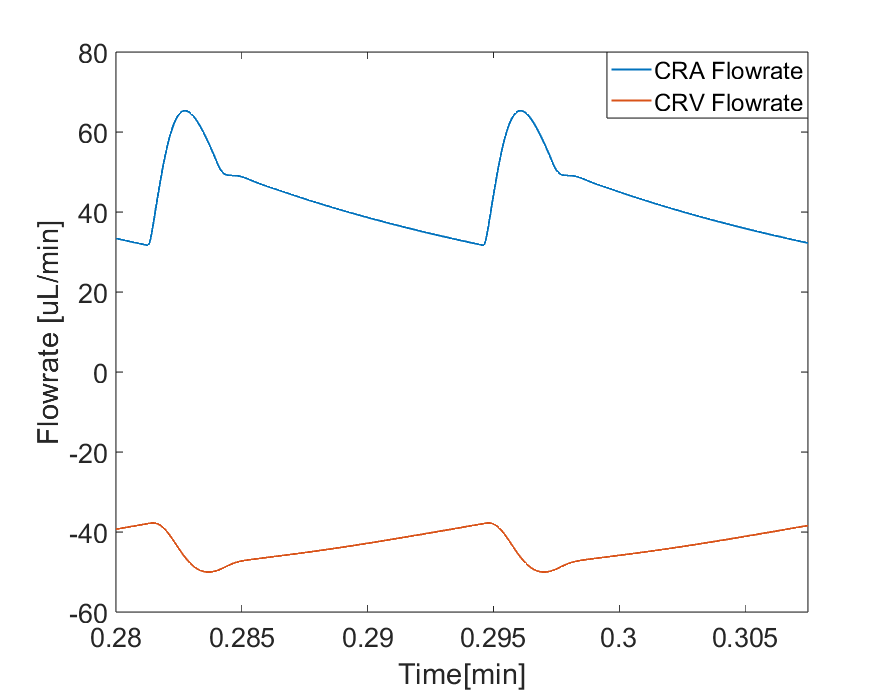}
		\caption{CRA and CRA blood flows.}
		\label{fig:validation:CRA_CRV_BF}
	\end{minipage} 
\end{figure}

The overall alignment of the model’s predictions with physiological trends is further illustrated in the Wiggers diagram (Fig. \ref{fig:validation:wiggers}), which provides a comprehensive visualization of pressure and volume dynamics throughout the cardiac cycle, reinforcing the physiological consistency of the simulated results on the cardiovascular side.

\medskip
\paragraph{The model predictions for key ocular biomarkers} are validated against quantitative indicators of blood flow in the central retinal artery (CRA) and central retinal vein (CRV) (Fig. \ref{fig:validation:CRA_CRV_BF}). 

Key metrics include:
\begin{itemize}
	\item CRA Mean Blood Flow (BF): average blood flow in the central retinal artery, compared to clinical measurements.
	\item CRA Peak Systolic and End-Diastolic BF: Maximum and minimum values of CRA BF.
	\item CRV Mean Blood Flow: Mean blood flow in the central retinal vein, considering different age groups.
	\item Mean Blood Pressures in the CRA ($P_{cra}$), retinal arterioles ($P_{art}$), retinal venules ($P_{ven}$) and CRV ($P_{crv}$).
\end{itemize}

\noindent
Table~\ref{tab:data:global} presents model predictions alongside experimental reference values. 

\begin{table}[ht]
	\centering
	\begin{tabular}{l l c c c}
		\hline \noalign{\smallskip}
		\textsc{Description} & \textsc{Unit} & \textsc{Value}  & \textsc{Reference} & \textsc{Present work}\\
		\hline \noalign{\smallskip}
		{CRA mean BF} & $[\mu l/min]$ & $40.91$ & \cite{guidoboni2014intraocular} & $ 46.56$ \\ 
		$ $          & $[\mu l/min]$ &  38.1 $\pm$ 9.1 & \cite{dorner2002calculation} \\
		&$[\mu l /min]$ &  33 $\pm$ 9.6 &  \cite{riva1979bidirectional}  \\[.08in]
		{CRA peak systolic BF} & $[\mu l /min]$ &120.6 $^*$ &  \cite{harris1996acute} &  129.46 \\
		& $[\mu l /min]$ & 122.5 $^*$ & \cite{julien2023} &  $ $ \\[.08in]
		{CRA end diastolic BF} & $[\mu l /min]$ & 30.1 $^*$ & \cite{harris1996acute} &  54.76 \\
		& $[\mu l /min]$ & 30 $^*$ & \cite{julien2023} &  $ $ \\[.08in]
		{CRV mean BF}  & $[\mu l /min]$&  64.9 $\pm$ 12.8  & \cite{garcia2002retinal}  & 43.47  \\
		\quad \textit{25 - 38 years} &  $[\mu l /min]$ &80 $\pm$ 12 &  \cite{feke1978laser} & $ $  \\
		\quad\textit{54 - 58 years} &  $[\mu l /min]$&73 $\pm$ 13 &   \cite{feke1978laser} & $ $\\[.08in]
		CRA Mean Pressure ($P_{cra}$) & [mmHg] & 43.92 & \cite{guidoboni2014intraocular} & 44.55 \\[.08in]
		Retinal Arterioles & [mmHg] & 36.09 & \cite{guidoboni2014intraocular} & 35.71\\
		Mean Pressure ($P_{art}$)\\[.08in]
		Retinal Venules Mean & [mmHg] & 22.13 & \cite{guidoboni2014intraocular} & 20.47\\
		Pressure ($P_{ven}$)\\[.08in]
		CRV Mean Pressure ($P_{crv}$) & [mmHg] & 18.84 & \cite{guidoboni2014intraocular} & 17.4\\[.05in]
		\hline \\[.08in]
	\end{tabular}
	\caption{Experimental data, blood flow. $^*$ uses assumption on CRA diameter value of about 160 $\mu m$ \cite{dorner2002calculation,lee2007association}.}
	\label{tab:data:global}
\end{table}

The model demonstrates a strong agreement with available clinical data for ocular circulation, particularly in predicting CRA and CRV blood flow. 
The mean CRA blood flow predicted by the model ($46.56 \: \mu l/min$) is consistent with experimental values reported in the literature, with minor deviations likely due to variations in assumed vessel diameters.

The peak systolic and end-diastolic blood flows in the CRA are also well reproduced, with the model predicting slightly higher values than some clinical measurements. This may stem from differences in experimental conditions or assumptions regarding systemic blood pressure fluctuations. Similarly, CRV blood flow values match well with reference data, reinforcing the model's ability to capture venous return dynamics.

Regarding pressures, these are not commonly available in standard clinical settings due to the difficulty of measurement. Therefore, we compared our predictions with those from a previously validated mathematical model \cite{guidoboni2014intraocular}, which is in agreement with our approach, to verify consistency.

In conclusion of this part, while minor deviations exist, the model successfully replicates cardiovascular and ocular circulation metrics, supporting its validity. Refinements in parameter estimation and integration of additional experimental data could further enhance predictive accuracy, particularly in capturing inter-individual variability.


\subsection{Predictive scenarios}
\label{subsec:prediction}

To investigate the effects of varying physiological parameters on both the cardiovascular and ocular systems, we conducted three simulations (A, B, and C), each focusing on different aspects of system dynamics.
The key parameters we considered for this prediction study were the left ventricle compliance (LVc) and the IOP.
Their values were adjusted to reflect physiological variations and assess their impact on the system at a global and a local level.

\subsubsection*{Simulation A: impact of an increase in IOP on the \eh model}
In this scenario, we investigated the effect of varying IOP from 15 mmHg to 30 mmHg.
This study is motivated by clinical insights highlighting the crucial role of venous circulation and the collapsibility of veins in ocular hemodynamics \cite{rai2024retinal}. 
Given the challenges associated with directly measuring venous parameters in clinical settings, mathematical modeling provides a valuable tool for inferring these values and gaining a deeper understanding of the underlying physiological mechanisms

\begin{figure}[ht]
	\begin{minipage}{0.49\linewidth}
		\centering
		\includegraphics[width=\linewidth]{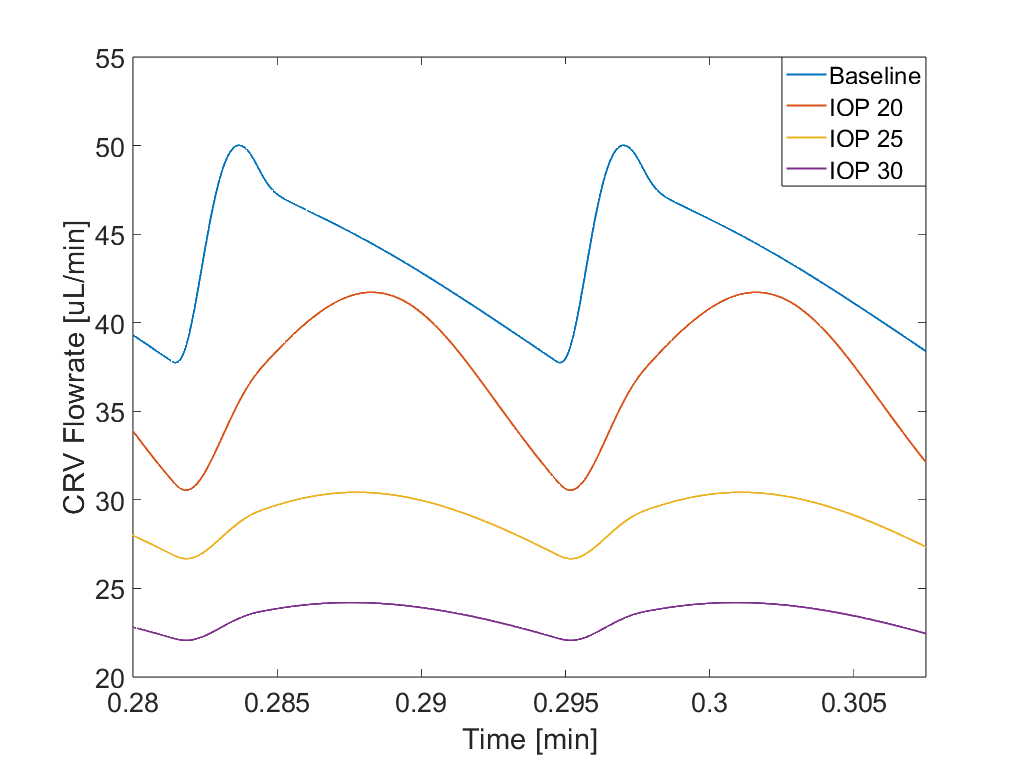}
		\caption{Scenario A: CRV blood flow. }
		\label{fig:scenarioA:CRV}
	\end{minipage} \hfill
	\begin{minipage}{0.49\linewidth}
		\centering
		\includegraphics[width=\linewidth]{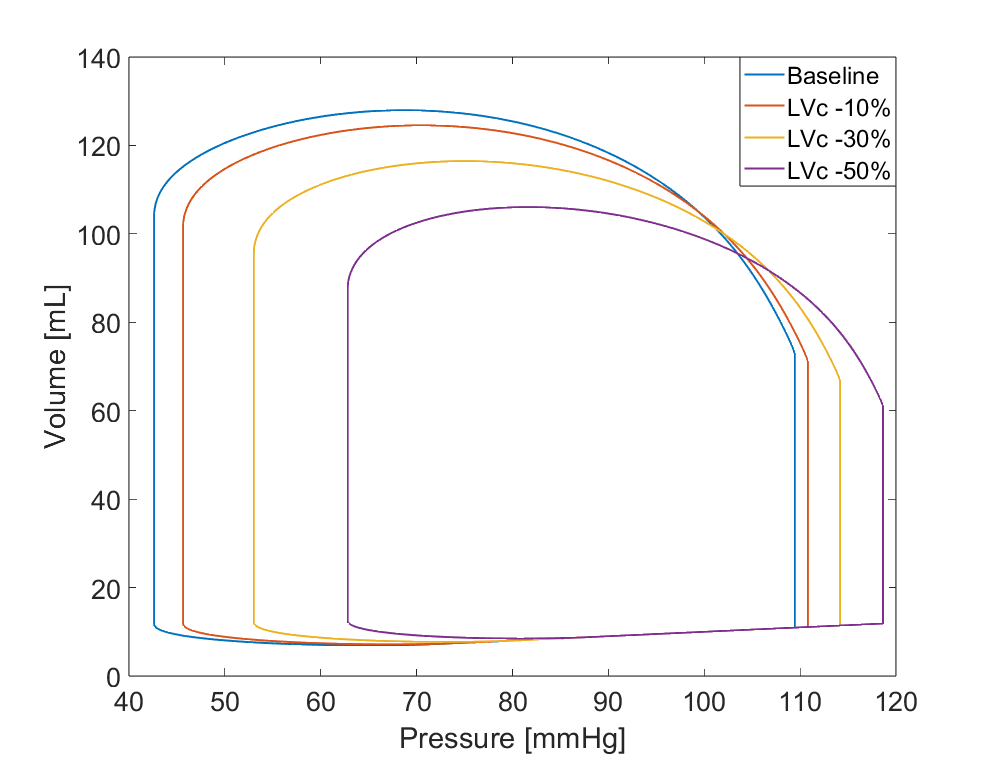}
		\caption{Scenario B: P-V loop }
		\label{fig:scenarioB:pv_loop}
	\end{minipage}
\end{figure}

As IOP increases, a marked decrease in CRA blood flow is observed, reflecting the restriction of the vascular supply due to elevated pressure. In contrast, CRV blood flow increases due to augmented resistance in the venous return pathway (see Fig. \ref{fig:scenarioA:CRV}).

The simulation results, reported in Table \ref{tab:res:A}, indicate a clear trend of decreasing CRA blood flow and increasing CRV blood flow as IOP rises. 
Despite these variations, the other cardiovascular parameters at a systemic level, such as SP/DP, EDV/ESV, and CO, remain largely unaffected by changes in IOP, as expected.

\begin{table}[ht]
	\centering
	\resizebox{\linewidth}{!}{
		\begin{tabular}{l l c c c c}
			\hline \noalign{\smallskip}
			\textsc{Output}& \textsc{Unit} & \textsc{IOP} = 15 mmHg & \textsc{IOP} = 20 mmHg & \textsc{IOP} = 25 mmHg & \textsc{IOP} = 30 mmHg \\
			\hline \noalign{\smallskip}
			SP / DP & [mmHg] & $128/69$ & $127/69$ & $127/69$ & $127/69$\\[0.08in]
			EDV / ESV & $[ml]$ & $113/43$ & $113/43$ & $113/43$ & $113/43$ \\[0.08in]
			CO & $[l/min]$ & $5.26$ & $5.26$ & $5.26$ & $5.26$\\[0.08in]
			CRA mean BF & $[\mu l /min]$ & $46.6$ & $43.6$ & $36.3$ & $30.8$ \\[0.08in]
			CRV mean BF & $[\mu l /min]$ & $43.5$ & $40.5$ & $29.1$ & $23.3$\\[.05in]
			\hline \\[.08in]
	\end{tabular}}
	\caption{Scenario A. IOP: Intraocular Pressure.}
	\label{tab:res:A}
\end{table}

\subsubsection*{Simulation B: effect of left ventricle compliance reduction on the \eh model.}

Simulation B focused on the impact of reducing LVc, adjusted by 10\%, 30\%, and 50\%. This change was modeled by altering the elastance scaling (ELS) parameter, which reflects the ability of the left ventricle to stretch and contract during the cardiac cycle.
This simulation setup builds upon the work of \cite{zaid2022mechanism}, which primarily investigated cardiac dynamics. 
Here, we extend the analysis to include ocular circulation, allowing for a comprehensive assessment of how changes in LVc influence both systemic and retinal hemodynamics.

The simulation results in Table \ref{tab:res:B} show a decrease in SP/DP and CO, as well as an increase in both EDV and ESV as LVc is reduced. 
These changes can be attributed to the decreased ability of the left ventricle to expand and contract effectively when its compliance is reduced. Lower LV compliance leads to less efficient filling and ejection, lowering pressures and cardiac output. At the same time, the reduced compliance causes the ventricle to hold more blood at both the end of diastole and systole, which is reflected in the increase in EDV and ESV. These results highlight how impaired ventricular compliance can significantly affect both the pumping efficiency and volume dynamics of the heart (see Fig. \ref{fig:scenarioB:pv_loop}).

On the ocular side, changes in LV compliance also impact the retinal blood flow. As LVc decreases, there is a reduction in CRA and CRV blood flows. These changes are more pronounced in CRA blood flow, which decreases in response to lower cardiac output. CRV flow is also affected, especially when LVc is reduced by 50\%. 
These results suggest that cardiovascular alterations may influence the ocular circulation, reflecting the dependencies between the heart and ocular dynamics.

\begin{table}[ht]
	\centering
	\begin{tabular}{l l c c c c}
		\hline \noalign{\smallskip}
		\textsc{Output}& \textsc{Unit} & \textsc{LVc: Baseline} & 
		\textsc{LVc: -10\% } & 
		\textsc{LVc: -30\% } &
		\textsc{LVc: -50\% } \\
		\hline \noalign{\smallskip}
		SP / DP & [mmHg] & $128/69$ & $123/69$ & $116/67$ & $107/61$ \\[0.08in]
		EDV / ESV & $[ml]$ & $113/43$ & $113/46$ & $114/53$ & $119/62$\\[0.08in]
		CO & $[l/min]$ & $5.26$ & $5.03$ & $4.59$ & $4.24$ \\[0.08in]
		CRA mean BF & $[\mu l /min]$ & $46.6$ & $45.6$ & $43.0$ & $38.1$\\[0.08in]
		CRV mean BF & $[\mu l /min]$ & $43.5$ & $43.3$ & $41.1$ & $35.2$\\[.05in]
		\hline \\[.08in]
	\end{tabular}
	\caption{Scenario B. \textsc{LVc}: Left Ventricle compliance.}
	\label{tab:res:B}
\end{table}

\subsubsection*{Simulation C: combined effects of LVc reduction and IOP increase}

Finally, Simulation C explores the combined effects of reducing LVc while simultaneously varying IOP.
Variations in IOP, similarly to Scenario A, affect the ocular system primarily at a local level, without significant impact on the overall cardiovascular dynamics.
As reported in Tables \ref{tab:res:C:sp_dp}, \ref{tab:res:C:edv_esv}, and \ref{tab:res:C:co}, the overall cardiovascular functions - SP/DP, EDV/ESV and CO, respectively - remain largely unchanged at varying IOP.
This confirms the hypothesis that IOP-related effects are confined to the local ocular circulation, while global system parameters remain stable.

\begin{table}[ht]
	\centering
	\begin{tabular}{l | c c c c}
		\hline \noalign{\smallskip}
		\bf \textsc{SP/DP} [mmHg] &
		\textsc{IOP} = 15 mmHg & \textsc{IOP} = 20 mmHg & \textsc{IOP} = 25 mmHg & \textsc{IOP} = 30 mmHg \\[0.05in]
		\hline \noalign{\smallskip}
		\textsc{LVc: Baseline} & $128/69$ & $127/69$ & $127/69$ & $127/69$ \\[0.08in]
		\textsc{LVc: -10\% } & $123/69$ & $123/69$ & $123/69$ & $123/69$ \\[0.08in]
		\textsc{LVc: -30\% } & $116/67$ & $116/67$ & $116/67$ & $116/67$\\[0.08in]
		\textsc{LVc: -50\% } & $107/61$ & $107/61$ & $107/61$ & $107/61$\\[0.05in]
		\hline \noalign{\smallskip}
	\end{tabular}
	\caption{Scenario C: systolic and diastolic blood central pressure (SP/DP).}
	\label{tab:res:C:sp_dp}
\end{table}

\begin{table}[ht]
	\centering
	\begin{tabular}{l | c c c c}
		\hline \noalign{\smallskip}
		\bf \textsc{EDV/ESV} [ml] &
		\textsc{IOP} = 15 mmHg & \textsc{IOP} = 20 mmHg & \textsc{IOP} = 25 mmHg & \textsc{IOP} = 30 mmHg \\[0.05in]
		\hline \noalign{\smallskip}
		\textsc{LVc: Baseline} & $112.8/42.6$ & $112.8/42.6$ & $112.8/42.6$ & $112.8/42.6$\\[0.08in]
		\textsc{LVc: -10\% } & $112.76/45.7$ & $112.8/45.7$ & $112.8/45.7$ & $112.8/45.7$ \\[0.08in]
		\textsc{LVc: -30\% } & $114.2/52.9$ & $114.2/52.9$ & $114.2/52.9$ & $114.2/52.9$\\[0.08in]
		\textsc{LVc: -50\% } & $118.6/62.1$ & $118.6/62.1$ & $118.6/62.1$ & $118.6/62.1$\\[0.05in]
		\hline \noalign{\smallskip}
	\end{tabular}
	\caption{Scenario C: left ventricle end diastolic volume (EDV) and end systolic volume (ESV).}
	\label{tab:res:C:edv_esv}
\end{table}

\begin{table}[ht]
	\centering
	\begin{tabular}{l | c c c c}
		\hline \noalign{\smallskip}
		\bf \textsc{CO} [l/min] &
		\textsc{IOP} = 15 mmHg & \textsc{IOP} = 20 mmHg & \textsc{IOP} = 25 mmHg & \textsc{IOP} = 30 mmHg \\[0.05in]
		\hline \noalign{\smallskip}
		\textsc{LVc: Baseline} & $5.26$ & $5.26$ & $5.26$ & $5.26$\\[0.08in]
		\textsc{LVc: -10\% } & $5.03$ & $5.03$ & $5.03$ & $5.03$ \\[0.08in]
		\textsc{LVc: -30\% } & $4.59$ & $4.59$ & $4.59$ & $4.59$\\[0.08in]
		\textsc{LVc: -50\% } & $4.24$ & $4.24$ & $4.24$ & $4.24$\\[0.05in]
		\hline \noalign{\smallskip}
	\end{tabular}
	\caption{Scenario C: cardiac output (CO).}
	\label{tab:res:C:co}
\end{table}

However, the combination of reduced LVc and increased IOP has a marked effect on ocular hemodynamics.
For the CRA (Tab. \ref{tab:res:C:craBF}), as IOP increases, blood flow decreases in a predictable manner, which is consistent with the vascular resistance caused by elevated pressure. However, the impact is moderate and does not lead to a dramatic alteration in flow until IOP reaches higher levels (\textit{e.g.}, 30 mmHg). This shows the resilience of the CRA in maintaining blood flow despite increasing IOP.

\begin{table}[ht]
	\centering
	\begin{tabular}{l | c c c c}
		\hline \noalign{\smallskip}
		\bf \textsc{CRA mean BF} $[\mu l /min]$ &
		\textsc{IOP} = 15 mmHg & \textsc{IOP} = 20 mmHg & \textsc{IOP} = 25 mmHg & \textsc{IOP} = 30 mmHg \\[0.05in]
		\hline \noalign{\smallskip}
		\textsc{LVc: Baseline} & $46.6$ & $43.6$ & $36.3$ & $30.8$\\[0.08in]
		\textsc{LVc: -10\% } & $45.6$ & $42.0$ & $35.0$ & $29.6$\\[0.08in]
		\textsc{LVc: -30\% } & $43.0 $ & $38.3$ & $31.8$ & $26.8$\\[0.08in]
		\textsc{LVc: -50\% } & $38.1$ & $33.6$ & $27.8$ & $23.1$\\[0.05in]
		\hline \noalign{\smallskip}
	\end{tabular}
	\caption{Scenario C: central retinal artery mean blood flow (CRA mean BF).}
	\label{tab:res:C:craBF}
\end{table}

More strikingly, the CRV (Tab. \ref{tab:res:C:crvBF} shows a pronounced sensitivity to both factors. As IOP rises, the pressure exerted on the veins increases, and this external pressure can interfere with venous return, especially when the internal pressure within the veins, driven by cardiovascular dynamics, is lower than the external pressure (IOP). This phenomenon, known as the Starling effect, can cause venous collapse and reduced blood flow.

For example, at IOP = 25 mmHg, the blood flow in the CRV is noticeably impaired even with baseline LVc. However, when LVc is reduced by 50\%, the CRV blood flow is significantly compromised even at lower IOP levels (\textit{e.g.}, IOP = 15 mmHg). This drop in blood flow in the CRV under lower IOP conditions suggests that the combination of reduced LVc and IOP increase could be indicative of certain pathologies, such as Normal Tension Glaucoma (NTG), where the vascular flow is impaired despite normal values of IOP.

\begin{table}[ht]
	\centering
	\begin{tabular}{l | c c c c}
		\hline \noalign{\smallskip}
		\bf \textsc{CRV mean BF} $[\mu l /min]$ &
		\textsc{IOP} = 15 mmHg & \textsc{IOP} = 20 mmHg & \textsc{IOP} = 25 mmHg & \textsc{IOP} = 30 mmHg \\[0.05in]
		\hline \noalign{\smallskip}
		\textsc{LVc: Baseline} & $43.5$ & $40.5$ & $29.1$ & $23.3$ \\[0.08in]
		\textsc{LVc: -10\% } & $43.3$ & $38.8$ & $28.5$ & $23.1$\\[0.08in]
		\textsc{LVc: -30\% } & $41.1$ & $35.0$ & $27.4$ & $22.4$ \\[0.08in]
		\textsc{LVc: -50\% } & $35.2$ & $31.7$ & $26.2$ & $21.2$\\[0.05in]
		\hline \noalign{\smallskip}
	\end{tabular}
	\caption{Scenario C: central retinal vein mean blood flow (CRV mean BF).}
	\label{tab:res:C:crvBF}
\end{table}

This shift in blood flow dynamics highlights the critical interaction between cardiovascular health and ocular pressure in regulating retinal blood flow. It also points to the potential for these combined factors to serve as biomarkers for ocular conditions like NTG, where blood supply to the retina may be compromised despite typical IOP values.

		\section{Discussions and conclusions}
		\label{sec:concl}

This study introduces the \eh model, a novel closed-loop framework designed to bridge the gap between cardiovascular and ocular dynamics. By integrating cardiovascular and retinal models, \eh provides a comprehensive platform for simulating the interconnected functions of these systems. 
Using a hydraulic-electrical analogy, the model effectively captures the dynamic interactions, offering a robust tool that can be adapted for multiple applications, such as studying the impact of cardiovascular diseases on ocular health and understanding how retinal pathologies are influenced by systemic circulation.

Section \ref{sec:results} presented the validation against clinical and experimental data demonstrating the model's ability to replicate key physiological parameters within acceptable ranges. 
For the cardiovascular system, parameters such as EDV, ESV, SV, CO, and EF align with clinical values, confirming the model’s capacity to simulate fundamental cardiac dynamics. Ocular parameters, including CRA and CRV blood flow, also align with experimental data, supporting the model's accuracy in retinal hemodynamics.

Additionally, scenario predictions demonstrate the model's ability to explore the effects of various physiological changes on both ocular and cardiovascular systems. For instance, simulations reveal how variations in LVc and IOP influence retinal blood flow and overall cardiovascular function. These predictions highlight the potential of the \eh model for detailed in-silico experimentation, allowing for the testing of different physiological states and their impact on ocular health and cardiovascular dynamics.

In terms of clinical applications, the \eh model holds significant potential for early detection of cardiovascular dysfunction, particularly through CRA waveform analysis. Previous work, such as that by the group \cite{sala2024analysis}, has demonstrated how the analysis of the CRA waveform can yield valuable insights into cardiovascular health. 
The integration of CRA waveform modeling within the \eh framework would allow for detailed and personalized simulation and analysis of how cardiovascular dysfunction, such as impaired heart pumping efficiency, can manifest in the retinal circulation.

Nevertheless, certain limitations are acknowledged. 
Discrepancies such as slight underestimations in EDV and ESV, as well as deviations in E$_{es}$ and E$_a$, likely stem from model simplifications, assumptions in pressure-volume relationships, and the exclusion of individual variability. 
Additionally, the "eye branch" simplification, while computationally efficient, does not fully represent the complexity of non-retinal ocular circulation and represents only a single eye. Furthermore, the use of 0D modeling and assumptions regarding parameterization, such as differences between males and females \cite{zaid2023cardiovascular}, introduces potential limitations in accuracy. 

Despite these limitations, the \eh model represents a significant advancement in the integrated modeling of cardiovascular and ocular systems. By providing a unified framework, this model offers a valuable tool for exploring the complex interactions between these systems and for investigating the potential impact of ocular dynamics on overall cardiovascular health. 
Future work should focus on refining model parameters, incorporating individual variability, and expanding the model to include more detailed representations of ocular substructures and both eyes. This expansion would enable the investigation of personalized treatments and their differential effects on each eye.
To better account for individual variability, sensitivity analysis and uncertainty quantification, as proposed in \cite{Prud_homme_2021}, should be incorporated.
Furthermore, extending the model to include a 4-chamber heart and both eyes would enhance its predictive capabilities, thereby improving its potential clinical applications for diagnosing and managing conditions that affect both the eye and the heart.

In conclusion, this study opens new avenues for experimental research investigating the relationship between patients' visual field deterioration and cardiac health. Understanding these connections could pave the way for new diagnostic approaches that link ocular hemodynamics with systemic cardiovascular conditions. This novel coupling of cardiovascular and retinal circulation models represents a significant step forward in exploring the interdisciplinary relationship between heart function and ocular health, with the potential to improve diagnostic and therapeutic strategies in both fields.

		\bibliographystyle{abbrv}
		\bibliography{biblio}
		
		\section*{Acknowledgments}
		Professor Giovanna Guidoboni has been partially supported
		by NSF DMS 2108711/2327640, NIH R01EY030851, NIH R01EY034718, and in part by a Challenge Grant award from Research to Prevent Blindness, NY. Professor Alon Harris is supported by NIH grants (R01EY030851 and R01EY034718), NYEE Foundation grants, The Glaucoma Foundation, and in part by a Challenge Grant award from Research to Prevent Blindness, NY. \\
		Professor Giovanna Guidoboni would like to disclose that she received remuneration
		from Qlaris and Foresite Healthcare serving as a consultant. All relationships listed
		above are pursuant to UMaine’s policy on outside activities. 
		Professor Alon Harris would like to disclose that he received remuneration from AdOM, Qlaris, and Cipla for serving as a consultant, and he serves on the board of AdOM, Qlaris, and SlitLed. Professor Alon Harris holds an ownership interest in AdOM, Oxymap, Qlaris, SlitLed, and AEYE Health. \\
		The authors are grateful to the American Institute of Mathematics for the support and hospitality during the SQuaRE program.

	\end{document}